\documentclass[prb,twocolumn,floatfix,amsfonts]{revtex4-1}
\usepackage{graphicx,graphics,color,epsfig}% Include figure files
\usepackage{bm}
\usepackage{amsmath}
\usepackage{amssymb}
\usepackage{epstopdf}
\usepackage{epsfig}
\usepackage{graphicx}

\usepackage{longtable}
\usepackage{rotating}
\usepackage{multirow}

\begin{document}
\title{Intrinsic interlayer electric field-induced switching regulatory mechanisms of photovoltaics and photocatalysis in Z-scheme MoSi$_{2}$N$_{4}$/WSi$_{2}$N$_{4}$ heterobilayers }

\author{Kai kong,\textit{$^{a}$} Qiang Wang,$^{\ast}$\textit{$^{a}$} Yan Liang,$^{\ast}$\textit{$^{b}$} Thomas Frauenheim,$^{\ast}$\textit{$^{c}$} Defeng Guo,$^{\ast}$\textit{$^{a}$} and Bin Wang$^{\ast}$\textit{$^{d}$}}

\address{$^a$ ~State Key Laboratory of Metastable Materials Science and Technology and Key Laboratory for Microstructural Material Physics of Hebei Province, School of Science, Yanshan University, Qinhuangdao 066004, People's Republic of China. E-mail: qiangwang@ysu.edu.cn; guodf@ysu.edu.cn}
\address{$^b$ ~College of Physics and Optoelectronic Engineering, Faculty of Information Science and Engineering, Ocean University of China, Songling Road 238, Qingdao 266100, People's Republic of China. E-mail: yliang.phy@ouc.edu.cn}
\address{$^c$ ~School of Science, Constructor University, 28759 Bremen, Germany; Computational Science and Applied Research Institute (CSAR), Shenzhen518110, P. R. China; Beijing Computational Science Research Center (CSRC), Beijing 100193, P. R. China. E-mail: thomas.frauenheim@bccms.uni-bremen.de}
\address{$^d$ ~Shenzhen Key Laboratory of Advanced Thin Films and Applications, College of Physics and Optoelectronic Engineering, Shenzhen University, Shenzhen, 518060, People's Republic of China. E-mail: binwang@szu.edu.cn}

\begin{abstract}
In the realm of modern materials science and advanced electronics, ferroelectric materials have emerged as a subject of great intrigue and significance, chiefly due to their remarkable property of reversible spontaneous polarization. This unique characteristic is not just an interesting physical phenomenon; it plays a pivotal role in revolutionizing multiple technological applications, especially in the domains of high-density data storage and the pursuit of fast device operation.
In the past few decades, there has been a significant increase in the number of investigations and commercial applications proposed for ferroelectric materials.
With the continuous miniaturization of electronic devices and the rapid development of two-dimensional (2D) materials, considerable efforts have been made towards exploring ferroelectricity in 2D materials, driven by the potential for revolutionary advancements in energy storage, data processing, and other emerging technologies. This exploration is fueled by the realization that 2D ferroelectric materials could offer unique properties such as high energy density, fast switching speeds, and scalability, which are crucial for the next generation of electronic devices.
The out-of-plane (OOP) ferroelectricity exhibited by these 2D materials is generally more advantageous than the in-plane ferroelectricity, primarily because the vertical polarizability aligns more seamlessly with the requirements of most practical technological applications. 
\end{abstract}

\maketitle

%\rule[-120pt]{3cm}{0.05em}

%* Corresponding authors.

%E-mail addresses:qingwang@ysu.edu.cn(Q.Wang),binwang@szu.edu.cn(B.Wang).

%\pacs{63.22.-m,65.80.CK, 72.80.Vp}
%\maketitle

\section{Introduction}
Recent growth in 2D materials offers new opportunities for ultrathin and  efficient solar photovoltaic systems, which can effectively utilize the solar energy.\cite{al2021solar,parida2011review} However, exploring novel 2D solar photovoltaic systems with superior performance is still in great demand to meet the huge energy consumption.\cite{allouhi2022up,tawalbeh2021} According to the different generation process of excitons (electron-hole pairs), the solar photovoltaic systems can be divided into two categories. One is based on the traditional bulk inorganic semiconductors, including Silicon, GaAs, and CdTe,\cite{LYJMCA4,2016large,papevz} in which the photoexcited carriers are generated without intermediate steps, thus lacking of long-range Coulomb interaction. Despite tremendous efforts, their widespread applications are still hindered by various significant challenges, such as the instability and the overly cumbersome production process of silicon crystal solar cells, the heavy metal elements contained in CdTe, and the undersized energy conversion efficiency of GaAs and so forth. Another is the excitonic solar cells (XSCs), which is based on the donor–acceptor composite networks, and the excitons are generated and dissociated concurrently at the donor–acceptor interface upon illumination.\cite{LYJMCA5,LYJMCA6} As indicated in previous studies, such excitons can be efficiently dissociated at the interfaces of XSC materials with different electron ionization and affinity potentials, thus substantial power conversion efficiencies (PCE) can be induced.\cite{ACS3} Even though, The PCE of most recent XSCs are still below 12\%, this is due to the fact that except for the promoting separation and hindering recombination of excitons, many other critical factors including effective visible light optical absorption, high carrier mobilities, small exciton binding energy, and moderate direct band gap ($1.2-1.6 eV$) of the donor, are also non-negligible.\cite{xu2020two,2018kagse} To this end, enormous challenges should still be overcome to design higher-performance XSCs, and searching new appropriate systems and exploring tunable mechanisms for improving the above critical factors still remain imperative.

Since the expeinmental mechanical exfoliation of graphene in 2004,\cite{gra2004} large amount of 2D materials have been reported in theory or laboratory one after another.\cite{2016h-BN,2018h-BN,2019MXenes,2017TMDS,2018TMDS} Due to the quantum confinement effect, 2D materials can exhibit more exotic properties and promising applications than their bulk counterparts,\cite{mak2010,20162d,gao20202d,zhang2020} making them more attractive for designing novel electronic and optoelectronic devices. Particularly, the recently reported new family of 2D ternary quintuple layers MoSi$_{2}$N$_{4}$/WSi$_{2}$N$_{4}$ have drawn tremendous interests for its thermoelectric and photovoltaic properties.\cite{mahmoud2019} Relevant researches and high-throughput calculations have demonstrated that the MoSi$_{2}$N$_{4}$/WSi$_{2}$N$_{4}$ monolayer can be easily experimental exfoliated from its bulk phase.\cite{2018kagse, xu2019new} Furthermore, a series of prominent photovoltaic behaviors have been verified in monolayer MoSi$_{2}$N$_{4}$/WSi$_{2}$N$_{4}$, including the moderate direct band gaps, high carrier mobilities, ideal visible light optical absorption coefficients, and remarkable performances in photovoltaic nano-devices.\cite{mahmoud2019} Even so, the invariable propertie of single 2D materials is far from enough to meet the urgent requirements of modern energy crisis. One effective way is constructing these existing 2D materials into heterojunctions,\cite{gobbi20182d,WQJMCA,su2020} which can enrich the uniqueness of these isolated components and break their limitations in different fields.\cite{yang2018,xu2020s} Especiaslly in 2D XSCs, due to the interlayers charge transfer and accumulation, enhanced photovoltaic characteristics of the heterojunctions ca

Here, based on the first-principles, the MoSi$_{2}$N$_{4}$/WSi$_{2}$N$_{4}$ vdWHs are constructed and studied systematicially. It is found that these vdWHs behave staggered type-II band alignment due to the charge redistribution induced by the unique interlayer stacking. Moreover, other numerous ideal characteristics are also found including the robust stabilitis, moderate direct band gaps, high carrier mobilities, efficient visible optical absorptions, considerable PCE and superior photocurrents in the 2D nano-devices, rendering them promising candidates for optoelectronic and photovoltaic devices. In light of the tunable interlayer charge transfer upon different vertical electric field ($E_{z}$), opposite band shifts occurs between the donor and acceptor layers. As a result, a phase transition from type-II to type-I band alignment is induced in such vdWHs, which may expand their applications in light-emitting diodes and lasers. More importantly, the above excellent photovoltaic characteristics of vdWHs could also be improved upon $E_{z}$, where the higher PCE, greater photocurrent and a red-shift peak of photocurrent in the visible light range can be obtained. These enhanced performances can further enrich their applications in photovoltaics, and the underlying mechanisms can also provide a theoretical guidance for futher experimental design and practical application of 2D photovoltaic systems, ecpecially for 2D XSCs. 
\begin{figure*}[!tb]
	\includegraphics[width=16cm]{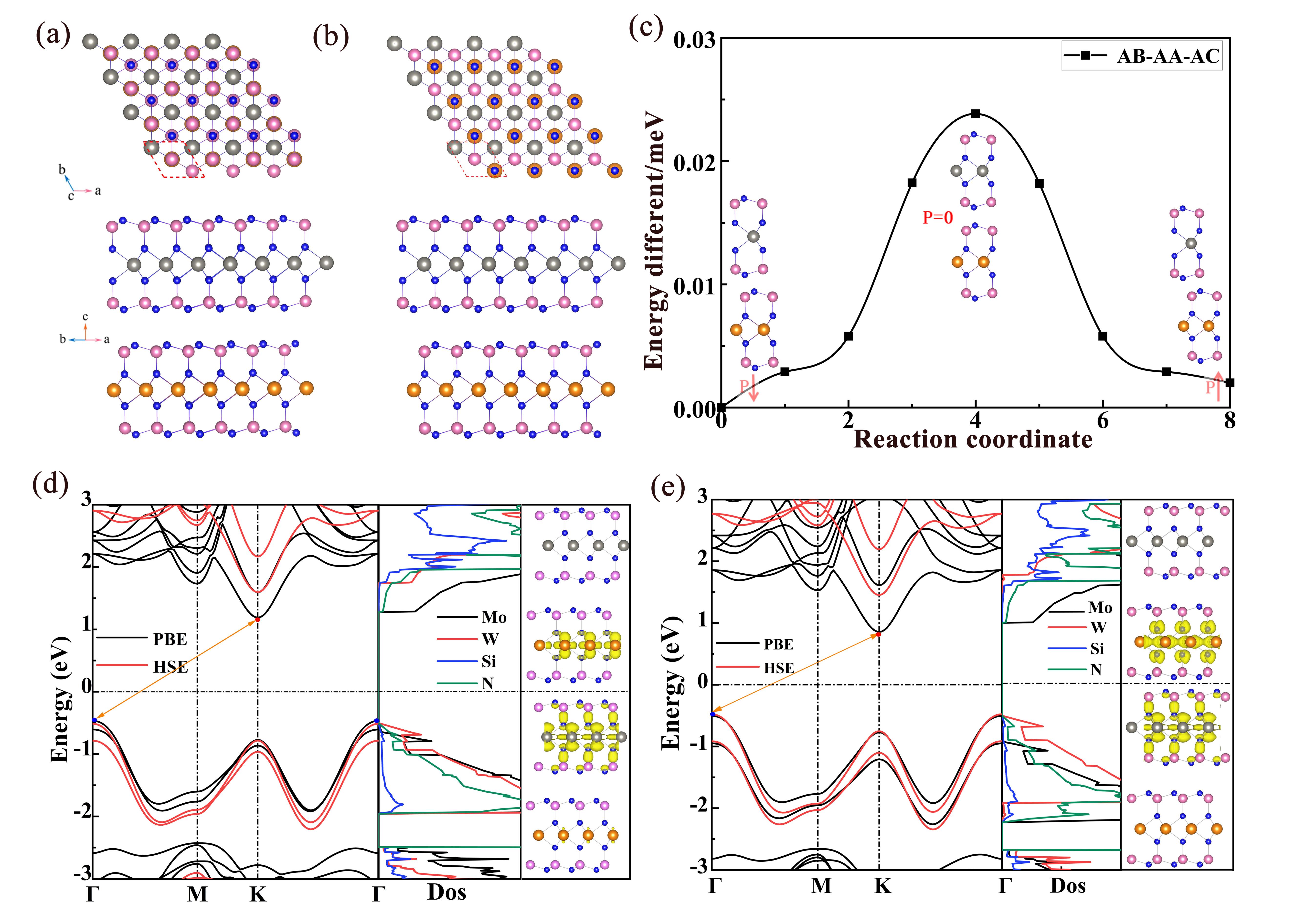}
	\centering
	\caption{ (a) and (b) Top and side views of the optimized crystal structures for AB- and AC-stacked MoSi$_{2}$N$_{4}$/WSi$_{2}$N$_{4}$ van der Waals heterostructures, respectively, with the unit cell outlined by a red diamond.(c) Energy barrier for ferroelectric switching in the MoSi$_{2}$N$_{4}$/WSi$_{2}$N$_{4}$ heterostructure, calculated using the nudged elastic band (NEB) method.
(d) and (e) The calculated band structure (left panel), partial density of states (PDOS, middle panel), and partial charge densities (right panel) at the conduction band minimum (CBM) and valence band maximum (VBM) for AB- and AC-stacked MoSi$_{2}$N$_{4}$/WSi$_{2}$N$_{4}$ heterostructures, respectively. The band structures are shown at the PBE and HSE06 levels, represented by black and red lines, respectively. The Fermi level is set to zero and is indicated by the green dotted horizontal line.}
	\label{fig1}
\end{figure*}
\section{Model and Numerical Method}
In this work, the structural and electronic properties of the VDW MoSi$_{2}$N$_{4}$/WSi$_{2}$N$_{4}$ were implemented by the Vienna ab initio simulation package (VASP) based on the density functional theory (DFT).
All first-principles calculations conducted in this study were performed using the Vienna Ab initio Simulation Package (VASP), in conjunction with the projector-augmented-wave (PAW) potential, which is rooted in the framework of density functional theory. For the purpose of geometrical optimizations, the exchange-correlation potential was approximated using the generalized gradient approximation (GGA) as formulated by the Perdew-Burke-Ernzerhof (PBE) functional.
The wave functions were formulated using the projector augmented wave (PAW) method employing a plane wave cut-off energy set at 400eV. To mitigate any spurious interactions perpendicular to the two-dimensional plane, a vacuum layer exceeding 20 Å was incorporated. For geometry optimization, a Monkhorst-Pack k-point grid of 11 × 11 × 1 was employed.

The structures were fully relaxed using convergence criteria of 0.01 eV/Å for forces and $10^{-6}$ 
 eV for energy. The phonon spectrum was generated using the Phonopy code. For the ab initio molecular dynamics simulation, a 3×3×1 supercell was employed with the canonical (NVT) ensemble at zero pressure and a temperature of 300K. The simulation ran for 5 picoseconds with a time step of 1 femtosecond. The energy barriers associated with ferroelectric switching were calculated using the nudged elastic band (NEB) method, while the ferroelectric polarization was evaluated through the Berry phase approach.

\section{Numerical Results and Discussions}

\begin{figure*}[!tb]
	\includegraphics[width=16cm]{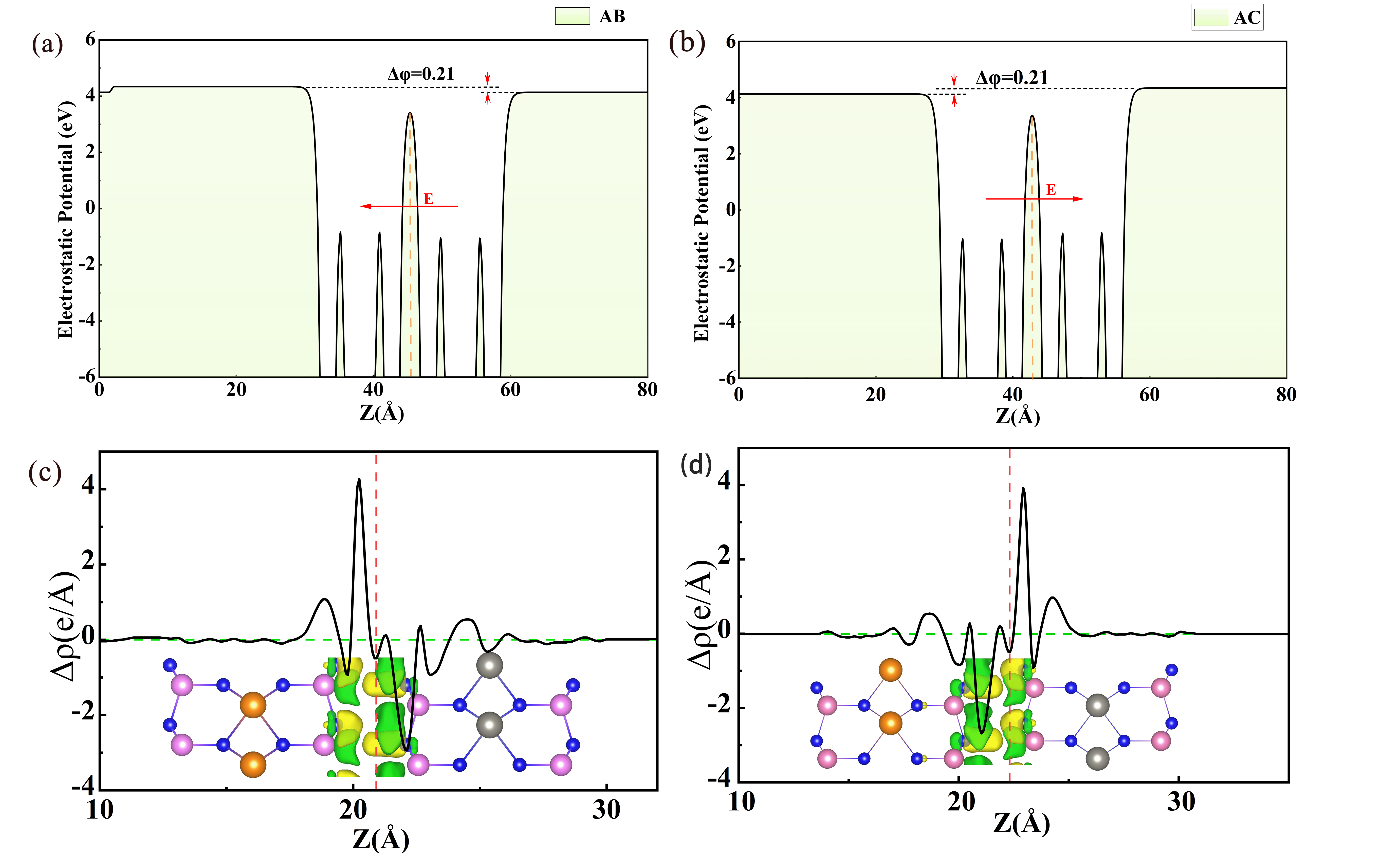}
	\centering
	\caption{ (a) and (b) Plane-averaged electrostatic potential differences along the vertical direction for AB- and AC-stacked MoSi$_{2}$N$_{4}$/WSi$_{2}$N$_{4}$ heterostructures, respectively.
(c) and (d) Charge density differences along the vertical direction for AB- and AC-stacked MoSi$_{2}$N$_{4}$/WSi$_{2}$N$_{4}$ heterostructures, respectively; yellow and cyan regions represent electron accumulation and depletion.}
	\label{fig2}
\end{figure*}
\subsection{Structures and stabilities of MoSi$_{2}$N$_{4}$/WSi$_{2}$N$_{4}$ vdWH} 
 
The 2D single layer of MoSi$_{2}$N$_{4}$ and WSi$_{2}$N$_{4}$(shown in Figure 1a) belongs to the MA$_{2}$Z$_{4}$ family, a class of semiconductors successfully synthesized by incorporating silicon during the chemical vapor deposition (CVD) growth process of molybdenum nitride, as demonstrated by Hong et al.31 This material consists of a Mo-N monolayer, which is sandwiched between two Si-N bilayers on the top and bottom. The atomic layers of MoSi$_{2}$N$_{4}$ are stacked in the sequence N-Si-N-Mo-N-Si-N, forming a layered structure. This arrangement contributes to the material’s outstanding structural stability and mechanical strength, making it suitable for various applications. The lattice parameter of the MoSi$_{2}$N$_{4}$ primitive cell, after full optimization, is found to be 2.91 Å, (meanwhile the WSi$_{2}$N$_{4}$ is found to be 2.93 Å) which is in excellent agreement with previous theoretical and experimental findings.Moreover, their monolayers have already been fabricated in experiments, demonstrating good dynamic and thermodynamic stability. To better tune their excellent properties in solar cells and photocatalysis, the MoSi$_{2}$N$_{4}$/WSi$_{2}$N$_{4}$ van der Waals heterostructures have been constructed,as shown in figure 2.
Here, we mainly consider three VDW stacking pattens:AA,AB and AC.Actually,the three pattens could be realized by the interlayer sliding, and the slip barrier is 0.025eV. Interesting,the phenomenon of interlayer sliding induces a polarization reversal both preceding and following the sliding . This can serve as an effective switch for manipulating the properties of heterojunctions.

Before exploring the electric properties of a heterostructure in greater detail, it is crucial to assess its stability to guarantee the practical viability of experiments.
Firstly, the structural stability was preliminarily verified by calculating the interfacial binding energy $E_{b}$. The formula is 
$E_{b}=(E_{vdw}-E_{MoSi_{2}N_{4}}-E_{WSi_{2}N_{4}})/S$, where the Evdw, Ebottom and Etop are per unit cell energies of the MoSi$_{2}$N$_{4}$/WSi$_{2}$N$_{4}$ vdWH, the free-standing MoSi$_{2}$N$_{4}$ and WSi$_{2}$N$_{4}$ monolayers, and S is the unit cell area.The calculated interfacial energy (Eb) for the MoSi$_{2}$N$_{4}$/WSi$_{2}$N$_{4}$ van der Waals heterostructure is XXX J/m², which is significantly higher than that of recently reported systems XXX.
As shown in table1.indicating a relatively stronger interlayer interaction in MoSi$_{2}$N$_{4}$/WSi$_{2}$N$_{4}$ vdWH.

Across the entire Brillouin zone of the phonon band dispersions, no imaginary frequencies are observed, suggesting that the structures attain stable minima on the potential energy surface.As a result, the MoSi$_{2}$N$_{4}$/WSi$_{2}$N$_{4}$ vdWH is dynamically stable. To further evaluate its thermal stability, we utilize ab initio molecular dynamics (MD) simulations to track the evolution of its total energy over time. Specifically, as depicted in Fig.3, we compute the total energy fluctuations of a 3 × 3 supercell at 300K, with a simulation duration of 1667 picoseconds and a time step of 3 femtoseconds. Our findings reveal that the total energy of MoSi$_{2}$N$_{4}$/WSi$_{2}$N$_{4}$ vdWH remains virtually unchanged throughout the simulation. Furthermore, the snapshot captured after relaxation clearly shows that the original structure is well-maintained, affirming the thermal stability of MoSi$_{2}$N$_{4}$/WSi$_{2}$N$_{4}$ vdWH at room temperature.
\subsection{Electronic character of MoSi$_{2}$N$_{4}$/WSi$_{2}$N$_{4}$ vdWH}

Generally, based on previous theoretical and experimental research, it has been confirmed that the obstructive influence of conventional type-II band alignment on exciton (electron-hole pair) recombination is advantageous for enhancing photocurrent generation.Meanwhile, the rate of electron hole recombination is also a key factor affecting photocatalytic water splitting. Here, it is worth noting that the breaking of interlayer symmetry leads to charge transfer, giving rise to the presence of interlayer built-in electric fields $(E_{\mathrm{int}})$. The electric fields further facilitate exciton separation. Depending on the staggered band arrangement and the direction of 
photogenerated carrier migration along the built-in electric field, the system can be categorized into either Z-type or II-type heterojunctions.Therefore, exploring this mechanism is of great significance in the fields of photovoltaic and photocatalysis.The charge density differences, denoted as $\Delta\rho(\textbf{r})$, are computed using the following equation:
\begin{equation}
	\Delta\rho(\textbf{r})
	 =[\rho_{MoSi_{2}N_{4}/WSi_{2}N_{4}}-\rho_{MoSi_{2}N_{4}}-\rho_{WSi_{2}N_{4}}]/n  
\end{equation}

\begin{figure*}[!tb]
	\includegraphics[width=15.5cm]{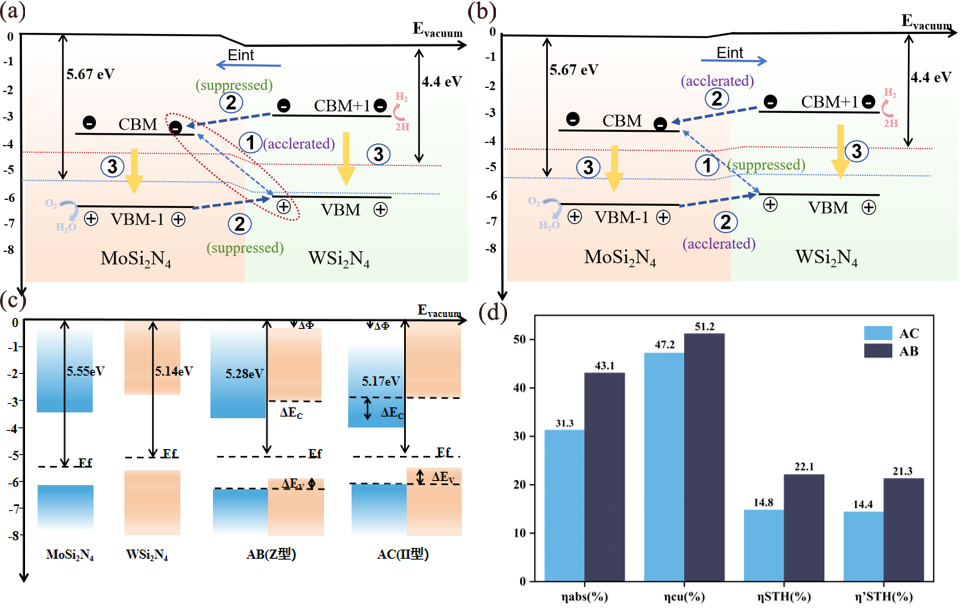}
	\centering
	\caption{(a) and (b) Band alignment with partial charge density of CBM (red) and VBM (blue) for AB-MoSi$_{2}$N$_{4}$/WSi$_{2}$N$_{4}$.The dotted lines are the standard redox potential levels of water splitting.(c) Schematic illustration of the band alignment between monolayer MoSi$_{2}$N$_{4}$ and WSi$_{2}$N$_{4}$ before and after the formation of the heterostructure.The work functions of each configuration are indicated, and the conduction band offset ($\Delta E_c$) and valence band offset ($\Delta E_v$) are also marked accordingly.(d) Solar-to-hydrogen (STH) efficiencies of the AB and AC configurations, along with the corresponding corrected values (STH$^\prime$) after incorporating the electrostatic potential adjustment.}
	\label{fig3} 
\end{figure*}

Here, $\rho_{MoSi_{2}N_{4}/WSi_{2}N_{4}}$, $\rho_{MoSi2N4}$, and $\rho_{WSi2N4}$ represent the charge densities of the MoSi$_{2}$N$_{4}$/WSi$_{2}$N$_{4}$ vdWH, the MoSi$_{2}$N$_{4}$ monolayer, and the WSi$_{2}$N$_{4}$ monolayer, respectively. The term n = 2 accounts for the two distinct components of the vdWH. Figure 2(a) presents both the 1D plane-integrated and 3D isosurface plots of $\Delta\rho(\textbf{r})$ at the interlayer region.Furthermore, according to the Bader analysis, a charge transfer of 0.002 electrons ($|e|$) occurs from the WSi$_{2}$N$_{4}$ layer to the upper MoSi$_{2}$N$_{4}$ layer.Consequently, this leads to the emergence of a net spontaneous electric polarization directed downwards during the process of such carrier separation in real space in
dicative for a direct Z-scheme vdWH.On the contrary, the charge transfer of the AC structure is opposite to that of AB, as shown in the right panel of Fig.~\ref{fig2} (a), where charges are transferred from the lower layer to the upper layer by a transfer amount of 0.0016 $|e|$, indicating that the direction of the built-in electric field is from bottom to top layer.Fig.~\ref{fig2} (b) illustrates the plane-averaged electrostatic potentials normal to the basal plane for these layers,The data results are consistent with the aforementioned statement.The value of discontinuity ($\Delta \Phi$) between the vacuum levels of the top and bottom layers of AB-MoSi$_{2}$N$_{4}$/WSi$_{2}$N$_{4}$ has been measured to be 0.21 eV, which is comparable or even larger than that of other polarized materials.(XXX)The electrostatic potential difference is primarily attributed to the separation of positive and negative charge centers, thereby breaking the spatial inversion symmetry.This asymmetry-induced interlayer built-in electric field is expected to significantly regulate electron-hole pair migration and recombination.
Interestingly, the polarization direction of the two structures can be reversed through interlayer sliding.

\begin{figure*}[!tb]
	\includegraphics[width=16cm]{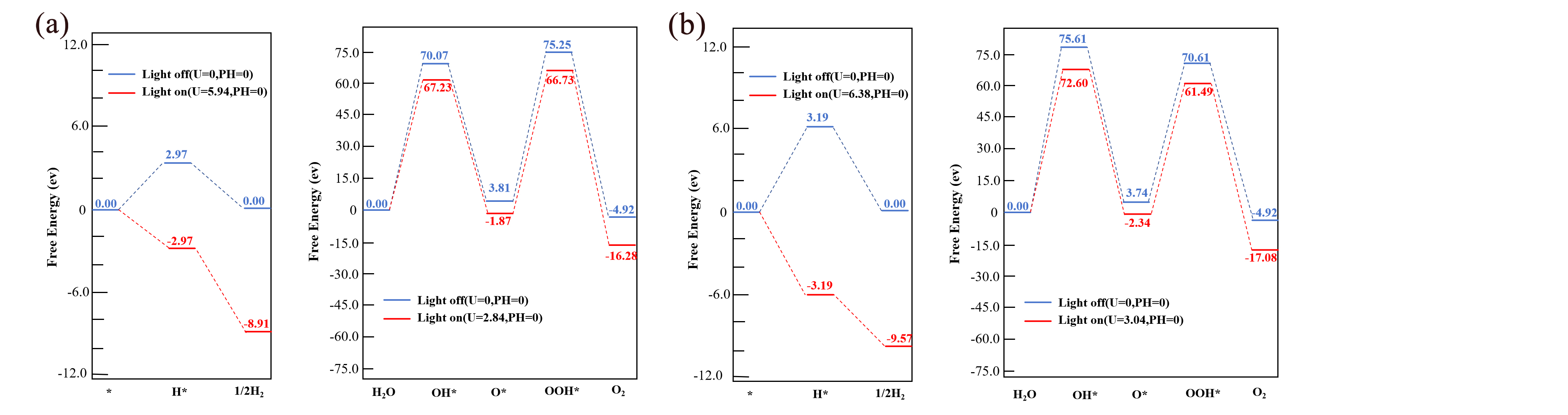}
	\centering
	\caption{ (a, b) Gibbs free energy diagrams of the hydrogen evolution reaction (HER) and oxygen evolution reaction (OER) on the AB- and AC-stacked MoSi$_2$N$_4$/WSi$_2$N$_4$ heterostructures, respectively.
} 
	\label{fig4}
\end{figure*}

Therefore, it is important to find out the pathway of exciton transfer.according to the schematic representation of the band edge alignment and the Z-scheme carrier-migration mechanism is illustrated in Fig.~\ref{fig2} (c). Under the classical mechanism, there are three transfer paths for charge carriers:Firstly,carrier recombination at the interface;Then,path 2 represents the transfer of interlayer electrons(holes)Paths 3 stands for the process of exciting transitions after electrons absorb photon energy (hv). Due to the presence of the built-in electric field, the recombination velocity of electron holes in path 1 is greater than that of electron and hole transfer in path 2 under the action of electric field force,suggesting a apparent Z-scheme path.In this scheme, a higher conduction band(CBM+1) and a lower valence band(VBM-1) are produced.Therefore, in the context of the MoSi$_{2}$N$_{4}$/WSi$_{2}$N$_{4}$ vdWH photocatalyst, the hydrogen evolution reaction (HER) and oxygen evolution reaction (OER) occur distinctly at the CBM+1 of W and VBM-1 of Mo, respectively.

\begin{figure*}[!tb]
	\includegraphics[width=15.5cm]{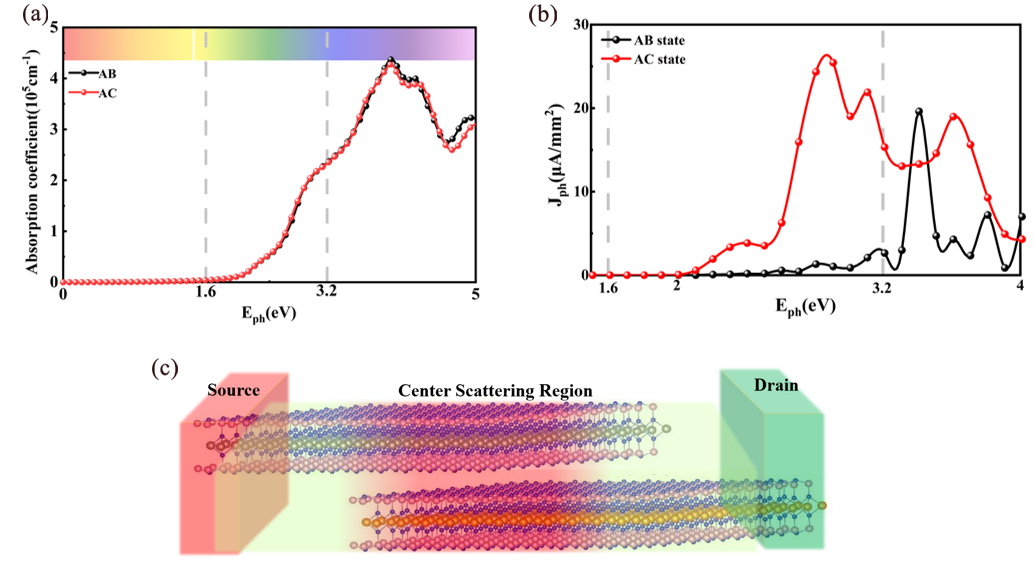}
	\centering
	\caption{ (a) Optical absorption coefficient $\alpha(\omega)$ as a function of incident photon energy ($E_{\mathrm{ph}}$) for the AB and AC configurations. The visible light region is indicated by the two vertical dashed lines.
(b) Photocurrent density ($J_{\mathrm{ph}}$) as a function of incident photon energy ($E_{\mathrm{ph}}$) at $\theta = 0^\circ$, under MoSi$_{2}$N$_{4}$/WSi$_{2}$N$_{4}$ vdWH states.
(c) Schematic illustration of a two-probe photovoltaic (PV) device based on the monolayer MoSi$_{2}$N$_{4}$/WSi$_{2}$N$_{4}$ vdWH.}
	\label{fig5} 
\end{figure*}
\subsection{Tunable effects of MoSi$_{2}$N$_{4}$/WSi$_{2}$N$_{4}$ vdWH on photo-systems}

As discussed above, Z-scheme band alignment difference from type-II band alignment is derived from the $E_{int}$ direction. 
Below, a meticulous examination and thorough analysis are conducted on the variations of key Photocatalytic(PC) and Photovoltaic(PV) characteristics of MoSi$_{2}$N$_{4}$/WSi$_{2}$N$_{4}$ vdWHs.For the optical harvesting abilities of this vdWH at different wavelengths are highly worthy of studying. It greatly depends on these crucial parameters such as the optical absorption coefficient ($a(\omega)$), the efficiency of photocatalysis ($\eta$),is significant.The value of $a(\omega)$ can be obtained using the equation provided below:
\begin{equation}
	a(\omega) = \frac{\sqrt{2}\omega}{c} \sqrt{\sqrt{\epsilon_1^2(\omega)+\epsilon_2^2(\omega)}-\epsilon_1(\omega)},
\end{equation}

Here, $\omega$ denotes the frequency of the incident light, and $\epsilon_1$ and $\epsilon_2$ represent the real and imaginary parts of the dielectric function, respectively. During the calculation process, $\epsilon_1$ is obtained through a summation over the empty states, whereas $\epsilon_2$ is computed using the Kramers–Kronig transformation.
As shown in Fig.~\ref{fig2} (a), when the incident light energy falls below the band gap, no light absorption occurs due to an insufficient driving force. However, once the incident light energy surpasses the band gap, vibration initiates and increases steadily with each increment in incident light energy. Within the visible range (1.6-3.2eV), the optical absorption coefficients for all three stacking modes continue to climb, with their peak absorption occurring in the ultraviolet region. Notably, the curves representing the AB and AC stacking modes in the three types are entirely superimposed above the curve for AA, a reflection of their mirror-symmetric spatial arrangement.In the entire light range, the maximum $a(\omega)$ of the MoSi$_{2}$N$_{4}$/WSi$_{2}$N$_{4}$ vdWH can reach $4.5 \times 10^5 \,\text{cm}^{-1}$ , higher than that of its individual components and surpasses that of other typical vdWHs including Sc2CO2/PtS2, KAgSe/KAgTe,etc., thus suggesting excellent visible light harvesting capabilities for the MoSi$_{2}$N$_{4}$/WSi$_{2}$N$_{4}$ vdWH. 

\subsection{Photocatalytic decomposition of water}

It is widely recognized that an ideal photocatalyst must meet two fundamental criteria. Firstly, it requires a suitable band gap to effectively harness incident visible or infrared light. Secondly, its band edge levels must straddle the redox potentials of the relevant reactants, ensuring adequate driving force for the the reaction to proceed.Based on the conventional reaction mechanism for water splitting, the band edge positions must align with the standard reduction potential of $\mathrm{H}^+$/H$_{2}$(-4.44 eV relative to the vacuum level) and the oxidation potential of O$_{2}$/H$_{2}$O(-5.67 eV relative to the vacuum level). In this work, we investigate the band edges of the VDW MoSi$_{2}$N$_{4}$/WSi$_{2}$N$_{4}$ utilizing the HSE06 computational level.As shown in Fig.~\ref{fig2} (d), it can be clearly seen that the band edge of type II heterojunction AC does not satisfy the necessary requirements for photocatalytic water decomposition. In contrast, the newly formed conduction band and valence band of the type Z heterojunction AB surpass both the reduction potential of hydrogen and the oxidation potential of oxygen, respectively.As a result, this type Z heterojunction AB can be effectively utilized as a catalyst in the process of photocatalytic water decomposition.
Therefore, in the following study, investigate the photoconductive (PC) properties of the Z-type heterojunction AB and the photovoltaic (PV) properties of the conventional type II heterojunction AC.
In fact, improving the efficiency of solar energy conversion is the ultimate goal of increasing the PC.Among them,the upper limits of light absorption ($\eta_{abs}$), carrier utilization $\eta_{cu}$ and solar-to-hydrogen efficiency ($\eta_{STH}$) are calculated shown as in Fig.~\ref{fig3} (b).
Here, $\eta_{STH}$ obeys the law of  $\eta _{STH}$ = $\eta _{abs}$ $\times$ $\eta _{cu}$,
Due to the unique carrier transfer mechanism of Z-type heterojunction,there is a significantly greater driving force to facilitate the redox reaction, resulting in a notably higher energy conversion efficiency. Specifically, the maximum efficiency achieved can reach XXX, surpassing the values reported in other studies. Consequently, this material holds promise as an excellent photocatalyst for water decomposition.

\subsection{Photocurrent in AB-MoSi$_{2}$N$_{4}$/WSi$_{2}$N$_{4}$-based nanodevices }

Due to the limitation of periodic boundary conditions, the photovoltaic properties of materials are currently only subject to qualitative predictions. Given that quantum scattering must be addressed under open boundary conditions in practical photoelectric nanodevices, the construction of MoSi$_{2}$N$_{4}$/WSi$_{2}$N$_{4}$ vdWH-based nanodevices is essential for investigating their actual photovoltaic performances.
Since the heterojunction has good spatial symmetry, resulting in little polarization between layers, we use a special method of constructing optoelectronic devices.Such structures are commonly known as "Staggered Stacking Heterojunction" or "Offset Stacked Structure." Their defining characteristic lies in the vertical transverse displacement between the two layers of two-dimensional material, meaning they are not perfectly aligned but exhibit a certain degree of dislocation. Due to the different band alignment in different regions, local built-in electric fields can be formed to optimize the electron-hole separation efficiency;The alteration of interfacial states resulting from dislocation modifies the electron transition pathways, thereby enhancing the utilization rate of photogenerated carriers;Since the upper and lower layers are not fully overlapping, the likelihood of direct recombination is decreased, leading to improved photoelectric conversion efficiency;The staggered arrangement facilitates an increased light absorption path, thus boosting the utilization rate of incident light in solar cells.
Fig.~\ref{fig5} (a) illustrate the schematic models of the two-probe PV nanodevice and the MoSi$_{2}$N$_{4}$/WSi$_{2}$N$_{4}$ vdWH-based two-probe photovoltaic device. In both designs, the entire scattering region is illuminated by vertically polarized light, with leads extending periodically from this region. These simplified device models offer several advantages compared to traditional metal-semiconductor tunneling junctions, including seamless and continuous interfaces, impeccable lattice matching (in the case of the MoSi2N4/WSi2N4 vdWH-based device), and straightforward fabrication. processes. Notably, the MoSi$_{2}$N$_{4}$/WSi$_{2}$N$_{4}$ vdWH-based device has been extensively employed in experimental battery design, underscoring its suitability and potential in this field.
During our simulations, a small bias voltage ($V_{bias}$ = 0.2 V) is applied between the source and drain, which is much smaller than the band gap of the vdWH and only used to drive photocurrent. When the photon energy ($E_{ph}$) exceeds the band gap, electrons are excited from the valence to the conduction band, generating photo-generated carriers. Then these carriers are separated in opposite directions under Vbias, finally producing the photocurrent. In the first order of the Born approximation, the photocurrent owing into the left probe can be expressed as:
\begin{equation}
	\begin{aligned}
		I_{L}^{ph}=\frac{ie}{h}\int \textbf{Tr}\left[\Gamma_{L}\{G^{<(ph)}+f_{L}(E)(G^{>(ph)}-G^{<(ph)})\}\right]dE,
		%J_{L}^{ph}=&\frac{ie}{h}\int \textbf{Tr}\left[\Gamma_{L}\{G^{<(ph)}+f_{L}(E)(G^{>(ph)}-G^{<(ph)})\}\right]dE,
	\end{aligned}  \label{cur0}
\end{equation}
where $\Gamma_L$ represents the line-width function, which signifies the coupling between the left lead and the central scattering region of the device. The greater/lesser Green's function of the system, denoted as $G^{>/<(ph)}$, encapsulates the electron-photon interaction, while $f_{L}$ denotes the Fermi distribution function associated with the left lead.
The calculated photocurrent density, $J_{ph}=I_L^{ph}/S$, of the MoSi$_{2}$N$_{4}$/WSi$_{2}$N$_{4}$ vdWH-based nanodevices at $E_{ph}$ equal to the corresponding band gaps is displayed in Fig.~\ref{fig6} (b). Due to the stronger built-in electriceld and higher absorption coefficient, a higher $J_{ph}$ can be detected for the device than for the MoSi$_{2}$N$_{4}$/WSi$_{2}$N$_{4}$ vdWH-based device.
All of this evidence demonstrates that 2D MoSi$_{2}$N$_{4}$/WSi$_{2}$N$_{4}$ vdWHs exhibit significant potential in the realm of 2D optoelectronic and photovoltaic applications. Furthermore, the physical mechanisms underlying the red-shift of the $J_{ph}$ peak in the visible-light region may provide valuable insights and open up new avenues for the design and manipulation of high-performance 2D photovoltaic devices in the future.

%\begin{figure*}[!tb]
%\includegraphics[width=16cm]{FigS5.jpg}
%\centering
%\caption{(color online) The density of states (Dos) versus energy of the monolayer BP (red curve) and bilayer 3R BP (green curves) structures under d=4.1{\AA}, 3.5{\AA}, and 2.9{\AA}, respectively. The green dash lines in each panel indicate the fermi levels.}
%\label{figS5}
%\end{figure*}

%and $\eta$ is defined by the ratio of maximum output photoelectron power $P_{out}$ density to the incident one $P_{in}$:
%\begin{equation}
%\eta=P_{out}^{max}/P_{in}
%\end{equation}
%Here, $P_{out}$ of the devices can be obtained by the product of $V_{d}$ and the corresponding $J_{ph}$.

{\section{Discussion}}
In summary, we have investigated the 
we have studied the high performances and feasibility of photovoltaic based on the MoSi$_{2}$N$_{4}$/WSi$_{2}$N$_{4}$ vdWHs. Their high mechanical, dynamic and thermal stability ensure high feasibilities of experimental synthesis. Their moderate direct band gaps, type-II band alignments, high carrier mobilities, efficient visible optical absorptions, ideal PCE and excellent photocurrent in the devices make them promising candidates for 2D high-performance photovoltaic devices. More interestingly, a phase transition of band alignment from type-II to type-I of the vdWH, combine with the further increased PCE (up to $23\%$) and a red-shift peak of the photocurrent in the visible light region are also detected under varying $E_{z}$, which not noly significantly improve and enrich the photovoltaic performances of these MoSi$_{2}$N$_{4}$/WSi$_{2}$N$_{4}$ vdWHs, but also shed light on the investigations on other similar 2D photovoltaic heterojunctions. Therefore, these calculations illustrate new avenues for next experimental design and manipulation of novel 2D solar photovoltaic systems.

\section*{Conflicts of interest}

The authors declare no competing financial interest.
%%%%%%%%%%%%%%%%%%%%%%%%%%%%%%%%%%%%%%%%%%%%%%%%%%%%%%%%%%%%%%%%%%%%%

\section*{Acknowledgements}
This work was financially supported by grants from the Natural Science Foundation of Hebei province (Grant No. E2019203163); Innovation Capability, Improvement Project of Hebei province (Grant No. 22567605H); Cultivation Project for Basic Research and Innovation of Yanshan University of China (Grant No. 2021LGQN017); Shenzhen Natural Science Foundations (Grant No. JCYJ20190808150409413), and the Young Talents Project at Ocean University of China.

%\bigskip
%\noindent{$^{\dag)}$ binwang@szu.edu.cn} \\
%\bigskip

%%%END OF MAIN TEXT%%%


%\balance


%%%REFERENCES%%%
\bibliography{main} %You need to replace "rsc" on this line with the name of your .bib file
\bibliographystyle{rsc} %the RSC's .bst file

\end{document}